\documentclass[aps,prb,twocolumn,superscriptaddress,notitlepage]{revtex4-1}
\usepackage{graphicx}
\usepackage[caption=false]{subfig}
\usepackage{float}
\usepackage{natbib}
\usepackage{xcolor}
\usepackage{xr}
\usepackage{amsmath}
\usepackage{amssymb}
\usepackage{array}
\usepackage{wasysym}
\usepackage{color,soul}
\usepackage{braket}
\usepackage{verbatim}
\usepackage{hyperref}
\usepackage{gensymb}
\usepackage{url}
\usepackage{dirtytalk}

\usepackage{filecontents}
\begin{document}

\title{Pressure induced transition from chiral charge order to time-reversal symmetry-breaking superconducting state in Nb-doped CsV$_3$Sb$_5$}

\author{J.N. Graham}
\thanks{These authors contributed equally to experiments.}
\affiliation{PSI Center for Neutron and Muon Sciences CNM, 5232 Villigen PSI, Switzerland}

\author{S.S. Islam}
\thanks{These authors contributed equally to experiments.}
\affiliation{PSI Center for Neutron and Muon Sciences CNM, 5232 Villigen PSI, Switzerland}

\author{V. Sazgari}
\affiliation{PSI Center for Neutron and Muon Sciences CNM, 5232 Villigen PSI, Switzerland}

\author{Y. Li}
\affiliation{Key Laboratory of Advanced Optoelectronic Quantum Architecture and Measurement, Ministry of Education, School of Physics, Beijing Institute of Technology, Beijing 100081, China}
\affiliation{Beijing Key Lab of Nanophotonics and Ultrafine Optoelectronic Systems, Beijing Institute of Technology, Beijing 100081, China}

\author{H.~Deng}
\affiliation{Department of Physics, Southern University of Science and Technology, Shenzhen, Guangdong, 518055, China}

\author{G. Janka}
\affiliation{PSI Center for Neutron and Muon Sciences CNM, 5232 Villigen PSI, Switzerland}

\author{Y. Zhong}
\affiliation{Institute for Solid States Physics, The University of Tokyo, Kashiwa, Japan}

\author{O. Gerguri}
\affiliation{PSI Center for Neutron and Muon Sciences CNM, 5232 Villigen PSI, Switzerland}

\author{P. Kr\'{a}l}
\affiliation{PSI Center for Neutron and Muon Sciences CNM, 5232 Villigen PSI, Switzerland}

\author{A. Doll}
\affiliation{PSI Center for Neutron and Muon Sciences CNM, 5232 Villigen PSI, Switzerland}

\author{I.~Bia\l{}o}
\affiliation{Physik-Institut, Universit\"{a}t Z\"{u}rich, Winterthurerstrasse 190, CH-8057 Z\"{u}rich, Switzerland}
\affiliation{AGH University of Science and Technology, Faculty of Physics and Applied Computer Science, 30-059 Krak\'{o}w, Poland}

\author{J.~Chang}
\affiliation{Physik-Institut, Universit\"{a}t Z\"{u}rich, Winterthurerstrasse 190, CH-8057 Z\"{u}rich, Switzerland}

\author{Z. Salman}
\affiliation{PSI Center for Neutron and Muon Sciences CNM, 5232 Villigen PSI, Switzerland}

\author{A. Suter}
\affiliation{PSI Center for Neutron and Muon Sciences CNM, 5232 Villigen PSI, Switzerland}

\author{T. Prokscha}
\affiliation{PSI Center for Neutron and Muon Sciences CNM, 5232 Villigen PSI, Switzerland}

\author{Y. Yao}
\affiliation{Key Laboratory of Advanced Optoelectronic Quantum Architecture and Measurement, Ministry of Education, School of Physics, Beijing Institute of Technology, Beijing 100081, China}
\affiliation{Beijing Key Lab of Nanophotonics and Ultrafine Optoelectronic Systems, Beijing Institute of Technology, Beijing 100081, China}
\affiliation{International Center for Quantum Materials, Beijing Institute of Technology, Zhuhai, 319000, China}

\author{K. Okazaki}
\affiliation{Institute for Solid States Physics, The University of Tokyo, Kashiwa, Japan}
\affiliation{Trans-scale Quantum Science Institute, The University of Tokyo, Tokyo, Japan}
\affiliation{Material Innovation Research Center, The University of Tokyo, Kashiwa, Japan}

\author{H. Luetkens}
\affiliation{PSI Center for Neutron and Muon Sciences CNM, 5232 Villigen PSI, Switzerland}

\author{R. Khasanov}
\affiliation{PSI Center for Neutron and Muon Sciences CNM, 5232 Villigen PSI, Switzerland}

\author{Z.~Wang}
\email{zhiweiwang@bit.edu.cn} 
\affiliation{Key Laboratory of Advanced Optoelectronic Quantum Architecture and Measurement, Ministry of Education, School of Physics, Beijing Institute of Technology, Beijing 100081, China}
\affiliation{Beijing Key Lab of Nanophotonics and Ultrafine Optoelectronic Systems, Beijing Institute of Technology, Beijing 100081, China}
\affiliation{International Center for Quantum Materials, Beijing Institute of Technology, Zhuhai, 319000, China}

\author{J.-X.~Yin}
\affiliation{Department of Physics, Southern University of Science and Technology, Shenzhen, Guangdong, 518055, China}

\author{Z. Guguchia}
\email{zurab.guguchia@psi.ch} 
\affiliation{PSI Center for Neutron and Muon Sciences CNM, 5232 Villigen PSI, Switzerland}

\date{\today}

\begin{abstract}
\textbf{The experimental realisation of unconventional superconductivity and charge order in kagome systems \textit{A}V$_3$Sb$_5$ is of critical importance. We conducted a highly systematic study of Cs(V$_{1-x}$Nb$_x$)$_3$Sb$_5$ with $x$=0.07 (Nb$_{0.07}$-CVS) by employing a unique combination of tuning parameters such as doping, hydrostatic pressure, magnetic fields, and depth, using muon spin rotation, AC susceptibility, and STM. We  uncovered tunable magnetism in the normal state of Nb$_{0.07}$-CVS, which transitions to a time-reversal symmetry (TRS) breaking superconducting state under pressure. Specifically, our findings reveal that the bulk of Nb$_{0.07}$-CVS (at depths ${\textgreater 20}$ nm from the surface) experiences TRS breaking below $T^*=40~$K, lower than the charge order onset temperature, $T_\mathrm{CO}$ = 58 K. However, near the surface (${\textless}$20 nm), the TRS breaking signal doubles and onsets at $T_\mathrm{CO}$, indicating that Nb-doping decouples TRS breaking from charge order in the bulk but synchronises them near the surface. Additionally, Nb-doping raises the superconducting critical temperature $T_\mathrm{C}$ from 2.5 K to 4.4 K. Applying hydrostatic pressure enhances both $T_\mathrm{C}$ and the superfluid density by a factor of two, with a critical pressure $p_\mathrm{cr}$ ${\simeq}$ 0.85 GPa, suggesting competition with charge order. Notably, above $p_\mathrm{cr}$, we observe nodeless electron pairing and weak internal fields below $T_\mathrm{C}$, indicating broken TRS in the superconducting state. Overall, these results demonstrate a highly unconventional normal state with a depth-tunable onset of TRS breaking at ambient pressure, a transition to TRS-breaking superconductivity under low hydrostatic pressure, and an unconventional scaling between $T_\mathrm{C}$ and the superfluid density.} 

\end{abstract}

\maketitle

\section{Introduction}

Exchange interactions in quantum materials are often balanced in such a way that there is significant overlap between the superconducting, magnetic and topological phases \cite{giustino20212021, neupert2022charge, khatua2023experimental}. However, if quantum materials are ever to mature into new technologies an important question to ask is can we stabilise one phase by tuning the exchange interactions through internal or external parameters. The kagome lattice---a network of corner sharing triangles---has been identified as a model system to explore such a question primarily due to its band structure which uniquely combines flat bands, van Hove singularities and topological Dirac points leading to a diverse set of quantum phases \cite{neupert2022charge, yin2022topological, guguchia2023unconventional, wang2023quantum, wilson2024v3sb5, deng2024chiral}. In this vein, the $A$V$_3$Sb$_5$ ($A = $K, Rb, Cs) compounds have been extensively studied and show several intriguing physical phenomena including giant anomalous Hall effect, pair density waves and high-temperature time-reversal symmetry (TRS) breaking charge order \cite{ortiz2019new, guguchia2023tunable, jiang2023kagome}. The physics of the \textit{A}V$_3$Sb$_5$ compounds is largely dictated by the competition or cooperation between two states, superconductivity and charge order. Extensive characterisation has shown that by suppressing the formation of charge order, the superfluid density and critical temperature, $T_\mathrm{C}$ can be increased, or vice versa. Depending on the system, previously tuning parameters have included external magnetic fields, hydrostatic pressure and depth dependent studies \cite{guguchia2023unconventional,guguchia2023tunable, gupta2022microscopic,graham2024depth}. 

\begin{figure*}
    \centering
    \includegraphics[width=\textwidth]{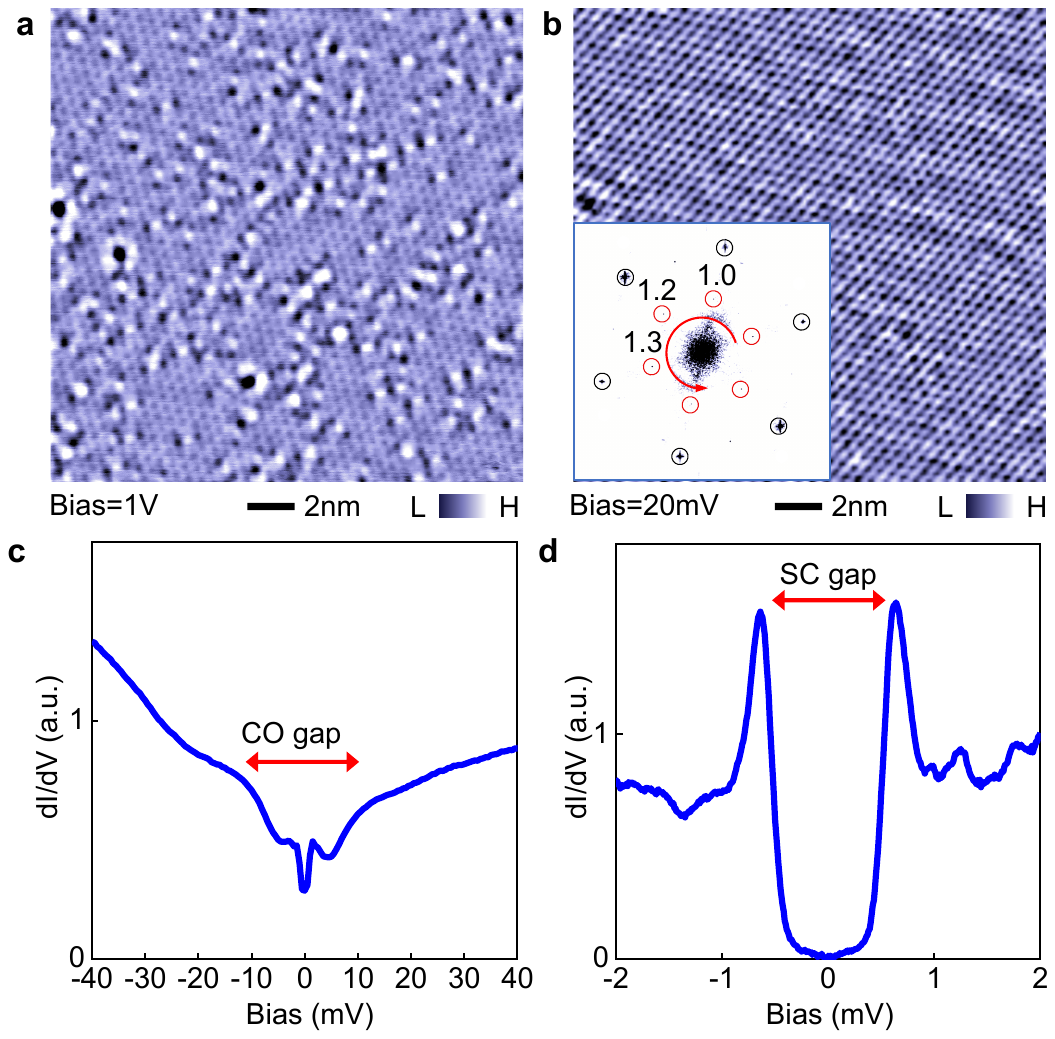}
    \vspace{-1.5cm}
    \caption{\textbf{Scanning tunnelling microscopy measurements  of Nb$_{0.07}$-CVS.} \textbf{a} Topographic image with high bias voltage showing the individual Nb dopants (dark spots). \textbf{b} Topographic image with low bias voltage at the same atomic location. The inset shows its Fourier transform, demonstrating the $2 \times 2$ charge order vector peaks (red circles). \textbf{c} Differential tunnelling spectrum showing a gap of $20~$meV caused by the charge order. \textbf{d} Differential tunnelling spectrum showing superconducting gap of $0.64~$meV. All the data were collected at a lattice temperature of 30mK. }
    \label{fig1}
\end{figure*}

Another tuning parameter is chemical doping, and lately it has been shown that these two key phases can be controlled by doping with small amounts of several elements, including Ta, As, Mo, Ti and Nb \cite{liu2022enhancement, li2022strong, liu2022evolution, liu2023doping, hou2023effect,zhong2023nodeless,  liu2024enhancement}. In each case, the dopant resides solely in the V-kagome layer, which suggests doping will always produce some effect on the magnetism \cite{mydosh2015spin, clark2021quantum, shi2024annealing}, however this effect is highly dependent on the dopant itself. For example, undoped CsV$_3$Sb$_5$ develops charge order at $T_\mathrm{CO} \approx 90~$K, and a superconducting transition at $T_\mathrm{C} = 2.5~$K \cite{ortiz2020cs}. Doping with $14~\%$ Ta \cite{zhong2023nodeless} or $4.7~\%$ Ti \cite{liu2023doping} was successful in fully suppressing the formation of the charge order after which the $T_\mathrm{C}$ of the Ta compound is increased to $5.2~$K. This is akin to applying hydrostatic pressure to undoped CsV$_3$Sb$_5$ \cite{gupta2022microscopic,zhu2022double,jiang2023kagome}. Meanwhile, doping with $7~\%$ Nb \cite{zhong2023nodeless} partially suppresses the formation of charge order from $90~$K to $58~$K, but raises $T_\mathrm{C}$ from $2.5~$K to $4.4~$K.

Inspired by these results, we have chosen to focus on one particular stoichiometry, Cs(V$_{0.93}$Nb$_{0.07}$)$_3$Sb$_5$ (Nb$_{0.07}$-CVS), which is the current limit of Nb doping into CsV$_3$Sb$_5$ \cite{zhong2023nodeless}. Recent ARPES results revealed Nb$_{0.07}$-CVS has a nodeless, nearly isotropic and orbital-independent superconducting gap \cite{zhong2023nodeless}, however there lacks a detailed microscopic investigation into the superfluid density and how we can manipulate the microscopic properties of both the superconducting and normal states of Nb$_{0.07}$-CVS which we aim to address in this article. In the following, we employ a unique combination of muon spin rotation/relaxation ($\mu$SR) techniques to investigate the temperature, field and depth dependence of the magnetic penetration depth and possible TRS breaking effects in the normal state. We also introduce another powerful tuning parameter, hydrostatic pressure, which we find has a strong influence on the superconducting properties of Nb$_{0.07}$-CVS, and we have constructed the phase diagram up to 2 GPa of applied pressure. Although the effect of these tuning parameters has been explored in other kagome systems, this is the first time all have been combined into a single study. This unique approach has revealed key features of the kagome system Nb$_{0.07}$-CVS. Our results show that Nb-doping decouples TRS breaking from charge order in the bulk, whilst aligning them near the surface. 
Under the application of low hydrostatic pressure (but still sufficient to suppress the charge order), the system transitions to a superconducting state with two notable unconventional features: (1) time-reversal symmetry breaking below $T_{\rm c}$ accompanied by nodeless electron pairing, pointing to chiral superconductivity, and (2) an unusual correlation between the superconducting critical temperature and the superfluid density.

\begin{figure*}
    \centering
    \includegraphics[]{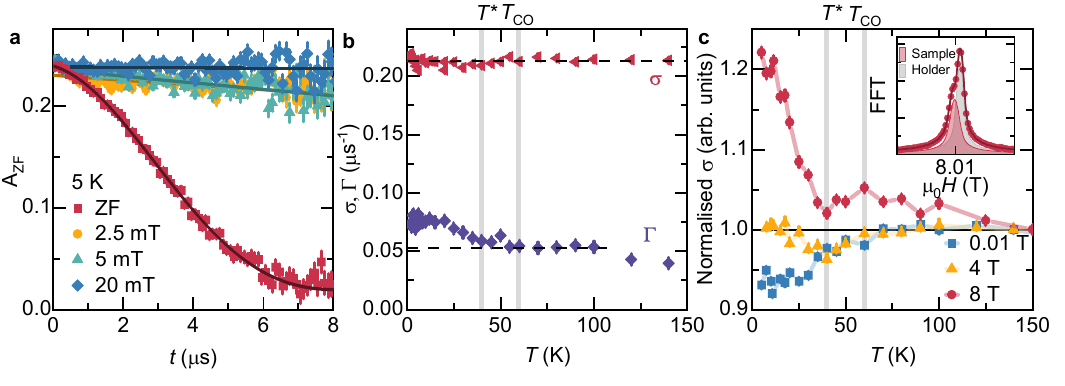}
    \caption{\textbf{Summary of $\mu$SR experiments in the normal state of Nb$_{0.07}$-CVS. a} Zero-field (ZF) and longitudinal-field (LF) muon-spin rotation spectra measured at $5~$K in various fields up to $20~$mT. \textbf{b} Evolution of $\sigma$ and $\Gamma$ relaxation rates as a function of temperature in the normal state. An increase in the $\Gamma$ relaxation below $T_\mathrm{CO}$ implies TRS breaking is present in Nb$_{0.07}$-CVS. \textbf{c} Evolution of $\sigma$ relaxation rate under applied fields of $0.01, 4$ and $8~$T. Inset shows the Fourier transform of the $8~$T data at $5~$K, which has been fit with two components; the sample (red) and the silver holder (grey).}
    \label{fig_ZF_LFl}
\end{figure*}

\section{Unconventional normal state}

Polycrystalline and single-crystal samples of Nb$_{0.07}$-CVS were synthesised according to other reports \cite{zhong2023nodeless}. Unless otherwise stated, measurements were performed on single-crystal samples. Firstly, Fig. 1 shows a summary of scanning tunnelling microscopy (STM) measurements that were collected at a lattice temperature of $30~$mK. Figure 1a shows individual Nb dopants that are atomically resolved and consistent with the target dopant level of $7~\%$. A lower bias voltage of $20~$meV, reveals that Nb$_{0.07}$-CVS is subject to a $2 \times 2$ chiral charge order, as indicated by the vector peaks (red circles) in the inset of Fig. 1b. This is the same charge order as the parent undoped compound, CsV$_3$Sb$_5$, however, the gap is reduced to $20~$meV (Fig. 1c), as compared to $40~$meV for CsV$_3$Sb$_5$ \cite{nakayama2021multiple}. Additionally, the STM results reveal an isotropic superconducting gap of 0.64 meV (Fig. 1d).

One of the benefits of $\mu$SR is the ability to detect internal magnetic fields as small as $0.01~$mT, without the need for the application of an external field, and is therefore an ideal tool to study the spontaneous fields that may arise due to TRS breaking in unconventional superconductors. Therefore, zero-field (ZF)-$\mu$SR measurements were conducted over a wide temperature range. A summary of the ZF-$\mu$SR measurements performed in the normal state of Nb$_{0.07}$-CVS are shown in Fig. \ref{fig_ZF_LFl}a and b and provide evidence of TRS breaking within the charge ordered state. Figure \ref{fig_ZF_LFl}a presents $\mu$SR measurements at $5~$K in ZF and longitudinal fields (LF - field is applied parallel to the muon spin polarisation) up to $20~$mT. To be consistent with previous works, the ZF data were fit with a Gaussian Kubo-Toyabe depolarisation function multiplied by an exponential term \cite{musrfit}:
\begin{equation}
    A_\mathrm{ZF}^\mathrm{GKT}(t) = \left(\frac{1}{3} + \frac{2}{3}(1 - \sigma^2t^2)\mathrm{exp}\left[-\frac{\sigma^2t^2}{2}\right]  \right)\mathrm{exp}(-\Gamma t) + A_\mathrm{bkg}
\end{equation}
where $\sigma/\gamma_\mu$ (${\gamma_{\mu}}$/2${\pi}$ = 135.5~MHz/T is the muon gyromagnetic ratio) is the width of the local field distribution primarily due to the nuclear moments, $\sigma$ and $\Gamma$ are muon spin relaxation rates, and $A_\mathrm{bkg}$ is a small, constant background term. The exponential relaxation rate $\Gamma$ can be due to a mixture of dilute and dense nuclear moments, the presence of electric field gradients or a contribution of electronic origin. The full polarization can be recovered by the application of a small external longitudinal magnetic field of $20~$mT. This highlights that the relaxation is due to spontaneous fields which are static on the microsecond timescale.

\begin{figure*}
    \centering
    \includegraphics[width = 18cm]{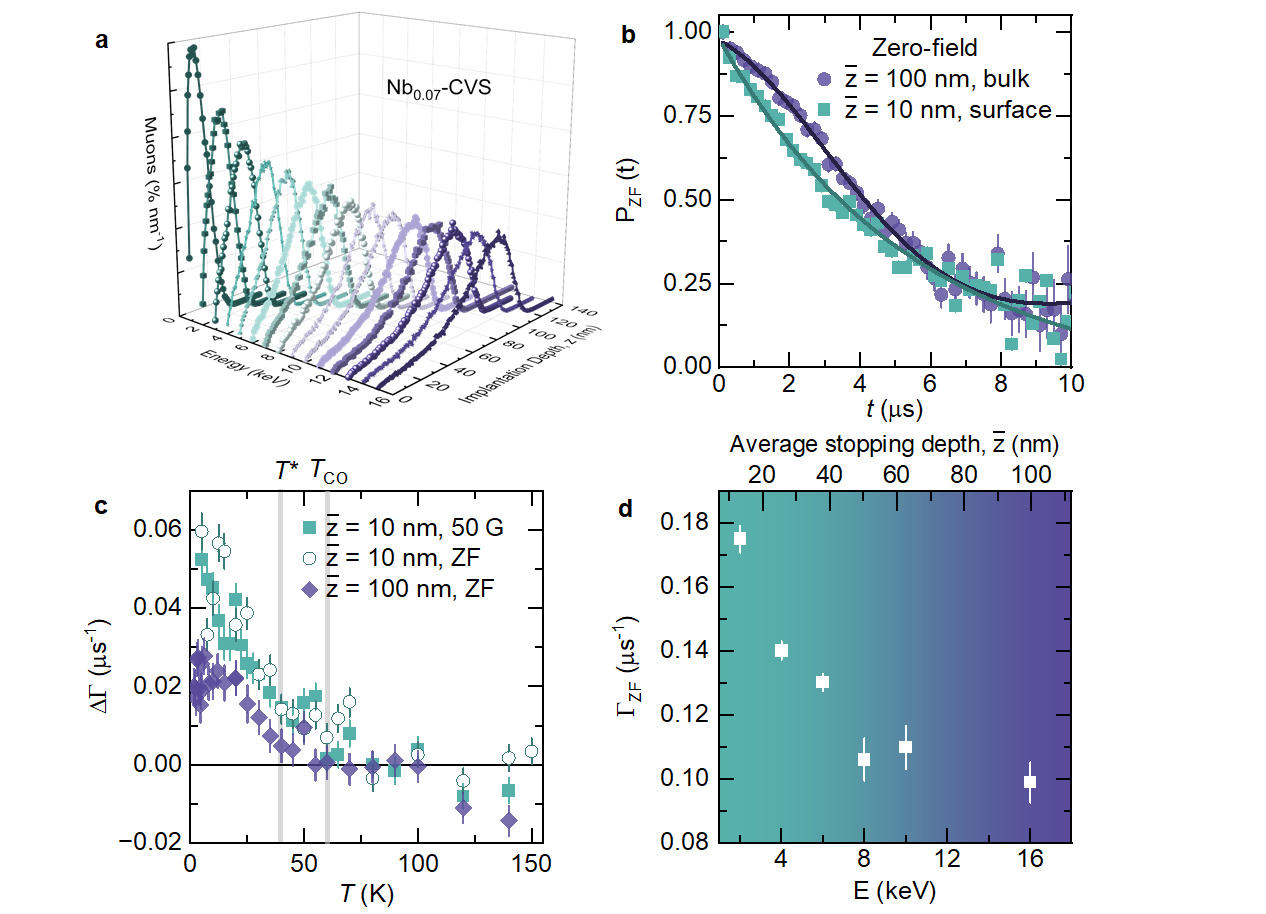}
    \caption{\textbf{Depth dependent magnetism in  Nb$_{0.07}$-CVS. a} Muon implantation profile simulated for several energies. \textbf{b} ZF $\mu$SR spectra for  Nb$_{0.07}$-CVS. obtained at $5~$K at the surface (mean implantation depth, $\bar{z} = 10~$nm) and in the bulk ($\bar{z} = 100~$nm). \textbf{c }Temperature dependence of the relaxation rate, $\Gamma$ measured in $50~$G and zero-field at the surface ( $\bar{z} = 10~$nm), and zero-field in the bulk  ($\bar{z} = 100~$nm). The high-temperature relaxation rate has been subtracted to put all data on a comparable scale. \textbf{d} The zero-field muon-spin relaxation rate measured at $5~$K in an applied field of $5~$mT as a function of muon implantation energy, $E$. Top axis shows the mean implantation depth, $\bar{z}$.}
    \label{fig3}
\end{figure*}

Figure \ref{fig_ZF_LFl}b shows the evolution of the $\sigma$ and $\Gamma$ rates across a wide temperature range measured in ZF conditions. Whilst the Gaussian $\sigma$ component remains nearly temperature-independent, displaying only a subtle, non-monotonic behaviour below $60~$K, a noticeable increase occurs below $T^* = 40$ K for the $\Gamma$ rate. It is noteworthy that this single-step increase of $\Gamma$ is different from the two-step transition which has been seen in $\mu$SR studies of undoped CsV$_3$Sb$_5$ \cite{khasanov2022time}, undoped RbV$_3$Sb$_5$ \cite{guguchia2023tunable,BonfaALC} and other kagome based compounds \cite{guguchia2023tunable, mielke2021nodeless}. The increase in the exponential relaxation below $T^{*}$ is estimated to be ${\simeq}$ 0.025~${\mu}$$s^{-1}$, which can be interpreted as a characteristic field strength ${\Gamma}$$_{12}$/${\gamma_{\mu}}$ ${\simeq}$ 0.05~mT. However, in the present case, these ZF-${\mu}$SR results alone do not provide conclusive evidence for time-reversal symmetry-breaking in Nb$_{0.07}$-CVS below $T^{*}$. The onset of charge order or charge redistribution might also alter the electric field gradient experienced by the nuclei and correspondingly the magnetic dipolar coupling of the muon to the nuclei. This can induce a change in the nuclear dipolar contribution to the ZF ${\mu}$SR signal. In order to substantiate the above ZF ${\mu}$SR results, systematic high field ${\mu}$SR \cite{guguchia2023tunable, mielke2021nodeless} experiments are therefore essential. In a strong magnetic field, the applied field determines the quantization axis for the nuclear moments, rendering the influence of charge order on the electric field gradient at the nuclear sites negligible.
 

To confirm the magnetic origin of the increase in $\Gamma$, we performed transverse-field (TF) $\mu$SR measurements under applied magnetic fields up to $8~$T, as shown in Fig. \ref{fig_ZF_LFl}c. The data at 0.01~T show a very weak temperature dependence similar to the temperature dependence of the zero-field nuclear rate ${\sigma}$ in Fig. \ref{fig_ZF_LFl}b. The results are normalised, so as to put the rates on a comparative scale. At $8~$T there is a large enhancement of the $\sigma$ relaxation rate below $T^* = 40~$K, and therefore confirms the magnetic origin of the TRS breaking. Relatively, the observed increase in $\sigma$ for Nb$_{0.07}$-CVS is larger than undoped CsV$_3$Sb$_5$ \cite{khasanov2022time}. Since the nuclear contribution to relaxation is unaffected by an external field, this implies that, at low temperatures and in high magnetic fields, the relaxation rate is predominantly governed by the electronic contribution. The inset shows the Fourier transform of the $8~$T data at $5~$K, where a clear splitting and broadening of the peak is visible. The narrow signal arises from the silver sample holder. The broad signal (red), with the fast relaxation, is shifted towards lower fields from the applied one, arises from the muons stopping in the sample and takes a majority fraction (70-80 ${\%}$) of the ${\mu}$SR signal. This points to the fact that the sample response arises from the bulk of the sample. The combined results of ZF-${\mu}$SR and high-field ${\mu}$SR reveal an enhanced internal field width below $T^{*} = 40$ K, providing direct evidence of time-reversal symmetry breaking fields within the kagome lattice. Notably, the onset of TRS breaking occurs at a lower temperature than the expected ordering temperature, $T_\mathrm{CO} = 58$ K. This indicates a splitting of these two temperatures in the bulk of the doped kagome system, a behaviour distinct from that observed in undoped compounds.

The measurements above are all presented from a bulk perspective. However, our recent studies have shown that TRS breaking is enhanced near the surface region of undoped RbV$_3$Sb$_5$ systems with optimal charge order \cite{graham2024depth}. We also demonstrated that in Ta-doped CsV$_3$Sb$_5$, where charge order is suppressed, no surface enhancement was observed. These findings therefore motivate a depth-dependent investigation of Nb$_{0.07}$-CVS, to explore whether the surface effect persists under the partial suppression of charge order. This is achieved through a combination of ZF and weak TF $\mu$SR measurements. Figure \ref{fig3}a shows the muon implantation profile for Nb$_{0.07}$-CVS for various implantation energies, simulated using the Monte Carlo algorithm TrimSP \cite{morenzoni2002implantation}. By varying the implantation depth of the muon we can therefore probe different depths of Nb$_{0.07}$-CVS. The ZF-$\mu$SR spectra are shown in Fig. \ref{fig3}b at both the surface ($E = 2~$keV, corresponding to a mean implantation depth, $\bar{z} = 10~$nm), and in the bulk ($E = 16~$keV, $\bar{z} = 100~$nm). A noticeable difference in the shape of field distribution is clearly visible between these two depths. Specifically, the ZF bulk data was analysed using Eq. 1, whereas the spectrum measured at the surface only required the exponential term, which is decoupled by applying a small external magnetic field longitudinally aligned with the muon spin polarization (B$_{LF}$ = 10 mT). This shows stronger static magnetism exists nearer the surface. Figure \ref{fig3}c presents the temperature dependence of the relaxation rate, $\Gamma$ in both ZF and TF measurements at the surface (green) and in the bulk (purple). To account for differences in background, this is shown in terms of the differences, $\Delta \Gamma = \Gamma(T) - \Gamma(150~\mathrm{K})$. In contrast to the bulk, at the surface the onset of TRS breaking is at $T_\mathrm{CO} = 58~$K indicating that charge order and TRS breaking are in sync at the surface but decoupled in the bulk. Additionally, the signal near the surface nearly doubles and continues to increase down to the lowest measured temperatures. Similar results were also observed in RbV$_3$Sb$_5$ \cite{graham2024depth}, however the response near to the surface was stronger by factor of 1.5 in RbV$_3$Sb$_5$ as compared to Nb$_{0.07}$-CVS. This suggests that the strength of the TRS breaking signal correlates with the onset/strength of charge order in these kagome superconductors. Finally, we measured the relaxation rate as a function of energy which is shown in Fig. \ref{fig3}d. These results highlight that the surface effects are confined to very near the surface with a characteristic depth of $\bar{z}_{c}$ ${\simeq}$ 20 nm, which is shorter than the study on RbV$_3$Sb$_5$  ($\bar{z}_{c}$ ${\simeq}$ 33~nm) \cite{graham2024depth}. This is significant, as it identifies a characteristic depth below which the materials properties differ from those of the bulk, which in the case of Nb$_{0.07}$-CVS is confined nearer to the surface.

\begin{figure*}
    \centering
    \includegraphics[]{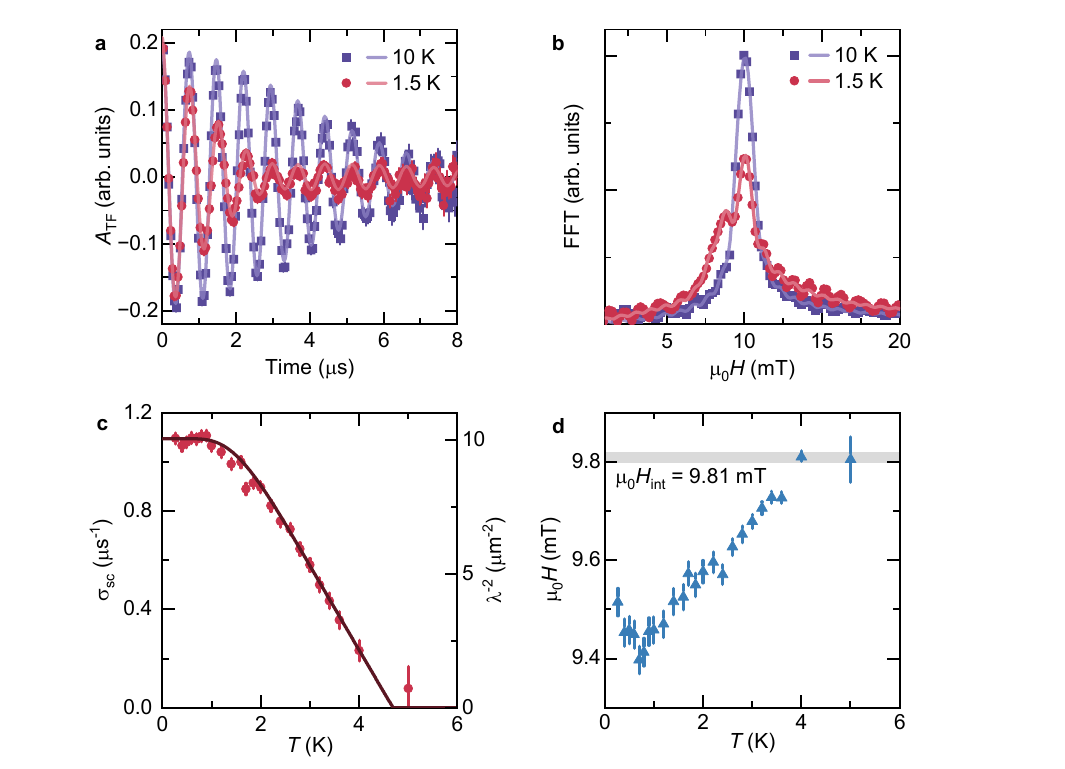}
    \caption{\textbf{Summary of $\mu$SR experiments in the superconducting state of Nb$_{0.07}$-CVS. a} Transverse-field (TF) $\mu$SR spectra collected above ($10~$K, purple) and below ($1.5~$K, red) $T_\mathrm{C}$ after field-cooling the sample from above $T_\mathrm{C}$ in an applied field of $10~$mT. \textbf{b} Fourier transforms of the data from \textbf{a}. \textbf{c} Temperature dependence of the superconducting muon spin depolarisation rate, $\sigma_\mathrm{sc}$ and inverse squared penetration depth, $\lambda^{-2}$ measured in $10~$mT. The solid line shows the fit of a nodeless $s$-wave gap model. \textbf{d} Response of the internal magnetic field, $\mu_0H_\mathrm{int}$ in the superconducting state.}
    \label{fig_TF}
\end{figure*}

\section{Unconventional superconducting state}

Next, we explored the microscopic superconducting properties of Nb$_{0.07}$-CVS at ambient and under hydrostatic pressures. A summary of the weak TF measurements in ambient conditions are shown in Fig. \ref{fig_TF}, which show that Nb$_{0.07}$-CVS has a bulk superconducting state, a dilute superfluid density and single nodeless superconducting gap structure.

Firstly, in Fig. \ref{fig_TF}a, measurements above (purple, $10~$K) and below (red, $1.5~$K) $T_\mathrm{C}$ show the expected response from a superconductor; a weakly relaxing oscillation above $T_\mathrm{C}$ due to the random local fields produced by the nuclear moments, which is strongly enhanced below $T_\mathrm{C}$ due to the formation of the flux line lattice (FLL). This is further evident in Fig. \ref{fig_TF}b, which shows the Fourier transform of the $\mu$SR spectra, which is sharp and symmetric above $T_\mathrm{C}$, but broadens and splits away from the applied field below $T_\mathrm{C}$. These data were analysed using the following functional form \cite{musrfit}:
\begin{equation}
    A_\mathrm{TF}(t) = \sum_i^n A_{S,i}e^ {\left[ - \frac{(\sigma_{\mathrm{tot},i}t)^2}{2} \right]}\mathrm{cos}(\gamma_\mu B_{\mathrm{int},i}t + \phi_i)
\end{equation}
where $A_\mathrm{S}$ is the initial asymmetry, $\sigma_\mathrm{tot}$ is the total muon spin depolarisation rate, $\gamma_\mu/(2\pi) \simeq 135.5~$MHz/T is the gyromagnetic ratio of the muon, $B_\mathrm{int}$ is the internal magnetic field, and $\phi$ is the initial phase shift of the muon ensemble. Data above $T_\mathrm{C}$ were fit with one component, whilst data below $T_\mathrm{C}$ were fit with three components; two originating entirely from the sample and making up $89.4\%$, with the third originating from the sample mount. Given that at least $89.4\%$ of the signal arises from the sample, and the diamagnetic response of the internal magnetic field (Fig. \ref{fig_TF}d) this confirms the bulk nature of superconductivity in Nb$_{0.07}$-CVS.

One of the most important measurable quantities in a $\mu$SR experiment on a superconuctor is the muon spin depolarisation rate, $\sigma_\mathrm{tot} (= \sqrt{ \sigma_\mathrm{sc}^2 + \sigma_\mathrm{nm}^2})$, which is comprised of superconducting, $\sigma_\mathrm{sc}$ and nuclear magnetic dipolar, $\sigma_\mathrm{nm}$ contributions, as $\sigma_\mathrm{sc}$ is directly proportional to the superfluid density, and as a result representative of the superconducting gap structure. The superconducting relaxation rate, $\sigma_\mathrm{sc}$ (Fig. \ref{fig_TF}c), was estimated by assuming the nuclear, $\sigma_\mathrm{nm}$ contribution was constant above $T_\mathrm{C}$ and subtracting this value in quadrature from $\sigma_\mathrm{tot}$. This measurement confirms $T_\mathrm{C} = 4.5~$K, and shows that like other kagome based superconductors, Nb$_{0.07}$-CVS has a dilute superfluid density, with the ratio $T_{\rm c}$/$\lambda_{\rm ab}^{-2}$ ${\simeq}$ 0.45 \cite{gupta2022microscopic,guguchia2023unconventional}---a hallmark feature of unconventional superconductivity. The topology of the superconducting gap structure is dependent on the temperature dependent behaviour of the superfluid density, and since the data plateau below $\sim~1~$K implies that a nodeless gap structure will be the most appropriate. This was confirmed through fitting the data, where it was found that in ambient conditions Nb$_{0.07}$-CVS has a $T_\mathrm{C} = 4.70(3)~$K, London penetration depth, $\lambda = 316(5)~$nm, and gap size, $\Delta = 0.590(5)~$meV. These results are in agreement with the STM results in Fig. 1d which show an isotropic gap of $0.64~$meV, and the previously performed ARPES measurements \cite{zhong2023nodeless}.

\begin{figure*}
    \centering
    \includegraphics[]{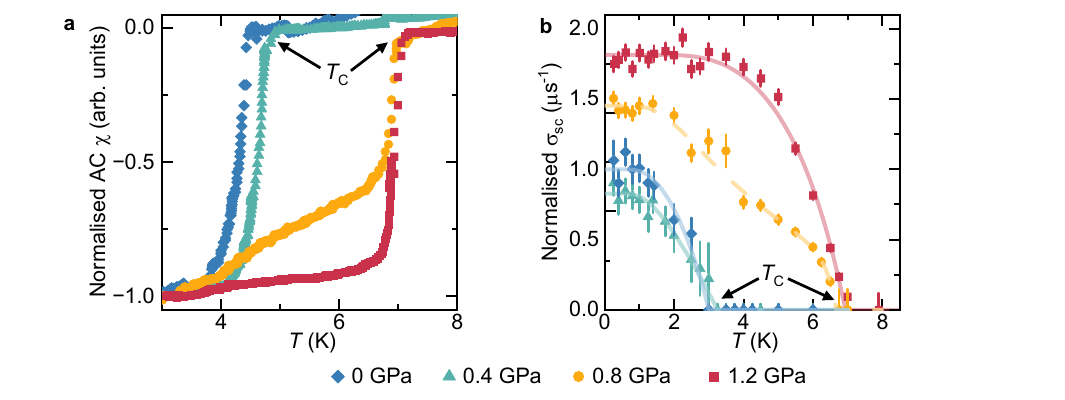}
    \caption{\textbf{Tuning superconductivity in Nb$_{0.07}$-CVS with pressure.} \textbf{a} AC magnetic susceptibility, $\chi$ under various applied hydrostatic pressures. Data is normalised between $-1$ and $0$. \textbf{b} Temperature dependence of normalised superconducting muon spin depolarisation rates, $\sigma_\mathrm{sc}$ measured in an applied field of $\mu_0H = 10~$mT at ambient and various applied hydrostatic pressures. The superconducting gap symmetries have been determined as single nodeless $s$-wave (solid lines) with the exception of $0.8~$GPa which has a double nodeless $s$-wave gap structure (dashed line). }
    \label{fig_musr_pressure}
\end{figure*}

An additional tuning parameter is pressure, which for kagome superconductors has a varied response, from none in LaRu$_3$Si$_2$ \cite{mielke2021nodeless} to significantly enhancing the superconducting properties in the $A$V$_3$Sb$_5$ compounds \cite{chen2021highly, gupta2022microscopic, hou2023effect}. Extending these investigations to include chemically doped samples is thus crucial to understand the mechanisms that underpin these behaviours. Figure \ref{fig_musr_pressure} displays an AC susceptibility and $\mu$SR study on the effect of hydrostatic pressure on Nb$_{0.07}$-CVS, which shows a strong response. In ambient and low applied pressures ($< 0.4~$GPa), the AC susceptibility confirms $T_\mathrm{C}$ as $\sim 4.5~$K. Increasing the pressure to $0.8~$GPa, leads to a jump in the $T_\mathrm{C}$ to $7~$K, but most intriguingly a change in shape of the superfluid density, and broadening of the AC data. These data were modelled with two nodeless gaps, as described in Table \ref{tab1}, and is similar to the superconducting gap model for undoped CsV$_3$Sb$_5$ at all pressures \cite{gupta2022microscopic}. Finally, at $1.2~$GPa, a smooth shape returns to the superfluid density, indicating a return to a single nodeless gap which is accompanied with a further increase in the superfluid density. It is notable that the agreement between the AC susceptibility and $\mu$SR data is better for higher pressures ($0.8$ and $1.2~$GPa) than the lower pressures (0 and $0.4~$GPa) which is due to $\mu$SR being a bulk probe so the transition to the superconducting state is usually only visible once the majority of the sample is superconducting (eg. at the tail end of the susceptibility curve). For the $1.2~$GPa data, it can be seen that the transition in the AC susceptibility is much sharper than the ambient pressure data, and therefore $T_\mathrm{C}$ is more in agreement between $\mu$SR and AC susceptiblity. Furthermore, the $\mu$SR data were collected under an applied field of $10~$mT whilst the AC data were collected under zero-field conditions, and this difference in experimental conditions can also result in some modification of $T_\mathrm{C}$. Table \ref{tab1} summarises the strong tuning ability of pressure in Nb$_{0.07}$-CVS, as between ambient and $1.2~$GPa, the superfluid density has doubled, the gap size has tripled from $\sim 0.5~$meV to $1.5~$meV, and there is a substantial increase in $T_\mathrm{C}$ from $4.5~$K to $7~$K.

\begin{figure*}
    \centering
    \includegraphics[]{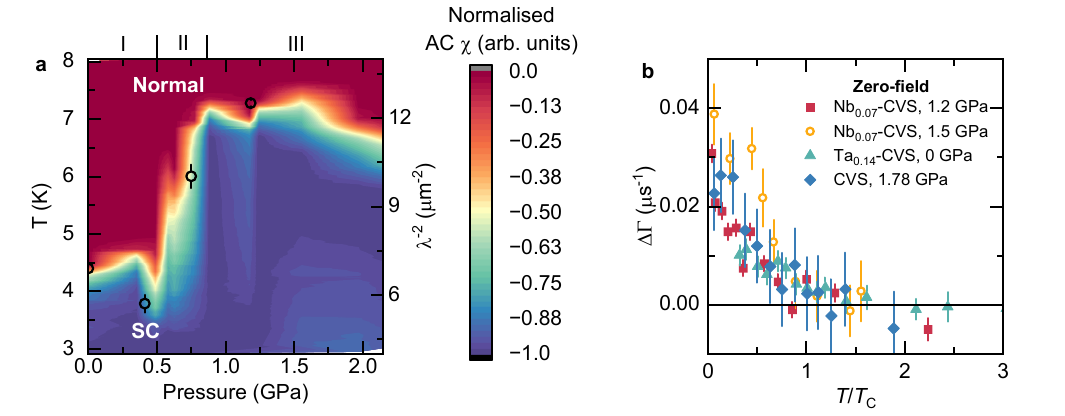}
    \caption{\textbf{Nb$_{0.07}$-CVS pressure phase diagram and comparison of TRS breaking in CsV$_3$Sb$_5$ derived compounds. a} AC susceptibility measurements as a function of applied hydrostatic pressure (0 - 2.2 GPa) have been normalised between -1 and 0 as described by the colour bar. The phase diagram is divided into three clear regions. Open circles show the inverse penetration depth squared, $\lambda^{-2}$ (right axis) determined from the gap structure analysis of the $\mu$SR data. \textbf{b} Comparison of TRS breaking across $T_\mathrm{C}$ for pure and doped CsV$_3$Sb$_5$ compounds. Single-crystal and polycrystalline samples are shown by closed and open markers, respectively. x-axis has been scaled by a factor of $T_\mathrm{C}$ for each compound at that pressure. }
    \label{fig_AC_pressure}
\end{figure*}

By systematically measuring more AC susceptibility data under pressure, we have been able to construct the pressure phase diagram of  Nb$_{0.07}$-CVS as shown in Fig. \ref{fig_AC_pressure}a (full data in Fig. \ref{Ext_data1} in Extended Data). This phase diagram shows the evolution of $T_\mathrm{C}$ from the measured AC susceptibility, normalised between -1 and 0 (as given by the colour bar) as a function of applied hydrostatic pressure. The transition from superconducting (blue) to normal (red) state can clearly be seen, with the broadness of the transition reflected in the width between the two. Here, we can clearly divide the behaviour of Nb$_{0.07}$-CVS into three regions: (I) the low pressure region - $p < 0.5~$GPa where there is a sharp transition at $T_\mathrm{C} = 4.5~$K, (II) a mixed region - $0.5 < p < 0.85~$GPa where the transition becomes very broad and nondescript, and (III) the high pressure region - $p > 0.85~$GPa where there is a sharp transition at $T_\mathrm{C} = 7 ~$K. These observations are quantitatively supported by the $\mu$SR measurements in Fig. \ref{fig_musr_pressure}b and Table \ref{tab1}, both in terms of $T_\mathrm{C}$ and the nature of the gap, which goes from single nodeless (I), to a double nodeless gap (II - reflecting the broadness of the transition), and back to a single nodeless gap (III). Also included in Fig. \ref{fig_AC_pressure} are the inverse penetration depths, $\lambda^{-2}$ (right axis, open circles) obtained from the gap structure analysis of the $\mu$SR data, which mirror the evolution of $T_\mathrm{C}$ very well and indicates nearly linear scaling between $T_\mathrm{C}$ and superfluid density. 

Next, it is crucial to investigate whether the superconducting state breaks time-reversal symmetry. To address this, we employed ZF-${\mu}$SR analysis. Since charge order already breaks TRS at a temperature $T^{\rm *} {\sim} T_{\rm CO} \gg T_c$, suppressing the charge order transition temperature ($T_{\rm CO}$) is essential, which may be achievable by applying pressure. At $0.8~$GPa, the pressure brings the system into the optimal $T_{\rm c}$ region of the phase diagram, where charge order is fully suppressed (see Fig. 6). In Fig. 6b, we present the internal field width $\Gamma$, derived from ZF-${\mu}$SR data, across the superconducting transition for Nb$_{0.07}$-CVS measured at 1.2 GPa and 1.5 GPa in single-crystal and polycrystalline samples, respectively. Notably, there is a significant enhancement in $\Gamma$, akin to that observed in superconductors that are believed to spontaneously break TRS, such as Sr$_2$RuO$_4$ \cite{LukeTRS}. A similar increase in $\Gamma$ below $T_{\rm C}$ is also observed for the undoped compound CsV$_3$Sb$_5$ at $p$ = 1.78 GPa and for Ta-doped CsV$_3$Sb$_5$, both exhibiting fully suppressed charge order \cite{gupta2022microscopic, graham2024depth}. This evidence strongly supports TRS-breaking superconducting states in Cs-derived kagome superconductors and aligns with findings in KV$_3$Sb$_5$ and RbV$_3$Sb$_5$, indicating an unconventional pairing state.

\begin{table}
    \centering
    \begin{tabular}{c|cccc}
    \hline
         Pressure (GPa)&  0&  0.4&  0.8& 1.2\\
         \hline
         $\lambda^{-2} (T = 0) (\mu$m$^{-2}$)&  6.9(3)&  5.7(3)&  10.0(4)& 12.5(1)\\
         $\lambda (T > 0)$ (nm)&  381&  419&  315& 283\\
         $\omega$&  1&  1&  0.73(4)& 1\\
         $T_\mathrm{C}$ (K)&  3.000(6)&  3.25(1)&  6.67(7)& 6.94(2)\\
         $\Delta_1$ (meV)&  0.54(9)&  0.47(7)&  0.65(5)& 1.56(4)\\
         $\Delta_2$ (meV)&  -&  -&  3.5(9)& -\\
         $\chi^2$&  4.5&  2.6&  15.4& 34.8\\
         $\chi^2$/NDF&  0.346&  0.153&  1.1& 1.74\\
         \hline
    \end{tabular}
    \caption{\textbf{Summary of superconducting gap structure parameters at different hydrostatic pressures}. $0, 0.4$ and $1.2~$GPa data were fit with a single nodeless $s$-wave gap, whereas the $0.8~$GPa data were fit with a double nodeless $s$-wave gap. $\lambda$ is the London penetration depth, $\omega$ is the phase fraction, $\Delta$ is the size of the gap, and $\chi^2$ and $\chi^2$/NDF are goodness of fit parameters.}
    \label{tab1}
\end{table}

\section{Discussion}

Our comprehensive investigation into the superconducting and normal-state properties of Nb$_{0.07}$-CVS has yielded three key findings:

(1) By combining zero-field and high-field muon spin rotation ($\mu$SR) measurements, we observed that time-reversal symmetry (TRS) breaking in the bulk of Nb$_{0.07}$-CVS (at depths greater than 20 nm from the surface) occurs below $T^* = 40$ K, which is lower than the charge order onset temperature, $T_\mathrm{CO} = 58$ K. In contrast, near the surface (within 20 nm), the TRS breaking signal is doubled in intensity and onsets at $T_\mathrm{CO}$, suggesting that Nb doping decouples TRS breaking from charge order in the bulk, while these phenomena remain synchronized at the surface. This behaviour contrasts with observations in the undoped CsV$_3$Sb$_5$ system, where we detected a two-step increase in the muon spin relaxation rate at $T_\mathrm{CO} = 90$ K and 30 K \cite{khasanov2022time,guguchia2023tunable}. In the undoped system, the two-step increase was attributed to the onset of TRS breaking associated with charge order, followed by an additional response, likely linked to a nematic state. In Nb$_{0.07}$-CVS, however, we observe a single TRS breaking transition. Regarding the enhanced TRS breaking signal near the surface in Nb$_{0.07}$-CVS, it mirrors what we previously noted in undoped RbV$_3$Sb$_5$ \cite{graham2024depth}, though in this case, the enhancement is two-fold rather than the five-fold increase seen in RbV$_3$Sb$_5$, which also displayed a more substantial rise in the onset temperature. In Ta-doped CsV$_3$Sb$_5$, where charge order is fully suppressed, no surface enhancement of TRS breaking was observed. In Nb$_{0.07}$-CVS, where charge order is partially suppressed, we observe a modest enhancement of TRS breaking near the surface. This suggests a correlation between the degree of surface enhancement and the strength of charge order.

(2) The superfluid density of Nb$_{0.07}$-CVS substantially increases with pressure. Within $1~$GPa of applied pressure, $T_\mathrm{C}$ increased from $4.5~$K to $7~$K, the superfluid density doubled and the gap size tripled. For this to occur, we assume that the charge order is suppressed with pressure, as has been measured in other $A$V$_3$Sb$_5$ compounds \cite{gupta2022microscopic, guguchia2023tunable}. The nearly linear scaling between the superconducting transition temperature ($T_\mathrm{C}$) and superfluid density is a hallmark of unconventional superconductivity. Combined with our previous observations of undoped vanadium-based kagome superconductors under pressure \cite{gupta2022microscopic, guguchia2023tunable}, this scaling suggests a generic characteristic of kagome superconductors and provides compelling evidence of unconventional superconductivity in these materials. Additionally, the sharp and well-defined transition from the low-$T_\mathrm{C}$ to high-$T_\mathrm{C}$ state strongly indicates a first-order phase transition in Nb$_{0.07}$-CVS. It is worth noting that the maximum $T_\mathrm{C}$ that is achieved in kagome materials, such as Ta-doped CsV$_3$Sb$_5$ \cite{YinNM}, is only once charge order is suppressed (with maximal chemical doping  $T_\mathrm{C}\sim5~$K). Hydrostatic pressure can also suppress charge order, but can elevate $T_\mathrm{C}$ to as high as 7–8 K in both undoped CsV$_3$Sb$_5$ and Nb$_{0.07}$-CVS. Unlike pressure, chemical doping typically introduces more disorder which suggests that disorder negatively impacts $T_\mathrm{C}$. Therefore, pressure emerges as a critical parameter for achieving the maximum possible $T_\mathrm{C}$ in these kagome materials, $A$V$_3$Sb$_5$.


(3) Nb$_{0.07}$-CVS exhibits a nodeless electron pairing, which is more pronounced under pressure as charge order is suppressed. Notably, ZF-$\mu$SR measurements below $T_\mathrm{C}$ at high pressure revealed an increase in relaxation rate, confirming that TRS is broken in the superconducting state. As illustrated in Figure \ref{fig_AC_pressure}b, this behaviour is not unique to Nb$_{0.07}$-CVS but aligns with findings across all CsV$_3$Sb$_5$-derived compounds examined by $\mu$SR \cite{graham2024depth, khasanov2022time}. Our results include compounds with suppressed charge order, such as Ta$_{0.14}$-CsV$_3$Sb$_5$ at ambient conditions \cite{graham2024depth} and undoped CsV$_3$Sb$5$ under 1.78 GPa pressure \cite{gupta2022microscopic, khasanov2022time}. Furthermore, we demonstrate that this TRS-breaking behaviour is consistent across both single-crystal (closed red squares) and polycrystalline (open yellow circles) samples of Nb$_{0.07}$-CVS, indicating that the mechanism underpinning the TRS breaking is robust to disorder from doping and hydrostatic pressure. The combination of a fully gapped pairing state and time-reversal symmetry breaking is compatible with chiral $d_{x^2-y^2} + id_{xy}$ or triplet chiral $p_x + ip_y$ states (where i represents the imaginary unit) \cite{PhysRevLett.127.177001, PhysRevB.86.121105}. Together with the dilute superfluid density and its nearly linear scaling with the critical temperature, these findings underscore the unconventional pairing in this system.


In conclusion, our work represents the most comprehensive microscopic study to date of both the normal and superconducting states in the doped kagome system Nb$_{0.07}$-CVS. We demonstrate how pressure drives the transition from a depth-tunable TRS-breaking normal state to a TRS-breaking superconducting state with significantly enhanced superfluid density and an increased superconducting critical temperature. This sheds light on the complex interplay between charge order, TRS breaking, and unconventional superconductivity in kagome materials. These findings highlight the pressing need for a comprehensive theoretical framework to uncover the fundamental mechanisms behind TRS breaking in superconductors, particularly given the unresolved nature of the order parameters in canonical TRS breaking superconductors like UTe$_{2}$ \cite{UTe2Science,UTe2JPSJ} and Sr$_{2}$RuO$_{4}$ \cite{SROmuSR,SROReview}.


\section*{Methods}

\subsection*{Muon spin rotation/relaxation ($\mu$SR)}
In a muon spin rotation/relaxation ($\mu$SR) experiment, a beam of nearly $100~\%$ spin polarised muons, $\mu^+$ are implanted into the sample, one muon at a time. These positively charged $\mu^+$ particles stop, due to thermal stabilisation, at interstitial lattice sites in the crystal. The muon will then precess at a frequency proportional to the local internal magnetic field, $B_\mu$, before decaying into a positron which is preferentially emitted in the direction of the muon spin, which is then detected.

\subsection*{Experimental details}

Zero-field (ZF), weak transverse-field (TF) and longitudinal-field (LF) $\mu$SR measurements, exploring both the superconducting and normal states were performed on the GPS and Dolly instruments \cite{amato2017new} at the Swiss Muon Source (S$\mu$S), Paul Scherrer Institute, Villigen, Switzerland. On GPS a continuous flow cryostat was used to measure temperatures in the range of $1.5~$K to $300~$K. Large single-crystals were placed in a mosaic arrangement, and the muon spin was rotated so that the measurements were sensitive to both the $ab$ plane and $c$-axis. A weak transverse field of $10~$mT was applied for measurements in the normal state. For LF measurements, a field of $2.5~$ to $20~$mT was applied, which was sufficient to fully decouple the muon spin from the internal magnetic field. Additional measurements were conducted on the high-field instrument, HAL also at S$\mu$S. Here, the applied field ranged from $10~$mT to $8~$T. Data in Fig. \ref{fig_ZF_LFl}c were fit with the following equation:
\begin{equation}
    A_\mathrm{TF}(t) = A_i \mathrm{exp}(-\sigma_i t)\mathrm{cos}\left(\gamma_\mu B_{\mathrm{int},i} t + \varphi\right)
\end{equation}
where $A_i$ is the asymmetry, $\sigma_i$ is the relaxation rate, $\gamma_\mu$ is the gyromagnetic ratio of the muon, $B_{\mathrm{int}, i}$ is the internal magnetic field and $\varphi$ is the phase shift. 

ZF and weak TF $\mu$SR experiments were conducted on single-crystal samples of Nb$_{0.07}$-CVS using the low energy $\mu$SR instrument LEM at the Swiss Muon Source (S$\mu$S), Paul Scherrer Institute, Villigen, Switzerland \cite{morenzoni2002implantation,prokscha2008new}. Large single-crystals, covering a total area of $2 \times 2~$cm$^{-2}$ were arranged in a mosaic layout on a nickel-coated plate and  adhered with silver paste. The sample was mounted on a cold finger cryostat, allowing temperatures of $5$-$300~$K to be accessed. The muon beam can be adjusted to energies between $1~$keV to $30~$keV. The implantation energy, $E$ corresponds to a specific muon implantation depth profile, allowing us to measure implantation depths between a few nanometres up to several tens of nanometres. In this experiment, the depths we probed were between $\bar{z} = 1 - 120~$nm. In both, Fig. \ref{fig3}a and d, we have provided a conversion between $E$ and $\bar{z}$. The muon implantation profiles in Nb$_{0.07}$-CVS for different implantation eneries were simulated using the TrimSP Monte Carlo algorithm \cite{morenzoni2002implantation}.

ZF and TF $\mu$SR experiments under pressure were performed at the high pressure spectrometer GPD \cite{khasanov2022perspective} at the Swiss Muon Source (S$\mu$S), Paul Scherrer Institute, Villigen, Switzerland. Randomly orientated single-crystal samples of Nb$_{0.07}$-CVS were loaded into a pressure cell in a compact cylindrical area of height $12~$mm and diameter of $6~$mm. The double wall pressure cell was made of an MP35N/CuBe alloy, specifically designed for high-pressure experiments and can reach applied pressures of $2.8~$GPa at room-temperature. Daphne 7373 oil was used a pressure medium to reach the highest pressures. The pressure was determined by tracking the superconducting transition of indium using AC magnetic susceptibility measurements. These AC measurements were also used to construct the pressure phase diagram using an in-house Janis cryostat which is controlled through the flow of liquid helium. Indium transitions were removed from the data for clarity. 

\subsection*{Superconducting gap structure}
To perform a quantitative analysis of $\mu$SR data and determine the superconducting gap structure, the superconducting muon spin depolarisation rate, $\sigma_\mathrm{sc}(T)$ in the presence of a perfect triangular vortex lattice is first related to the London penetration depth, $\lambda(T)$ by the following equation \cite{London_muSR, London_muSR2}:
\begin{equation}
    \frac{\sigma_\mathrm{sc}(T)}{\gamma_\mu} = 0.06091 \frac{\Phi_0}{\lambda^2(T)}
\end{equation}
where $\Phi_0 = 2.068 \times 10^{15}~$Wb is the magnetic flux quantum. This equation is only applicable when the separation between vortices is larger than $\lambda$. In this particular case, as per the London model, $\sigma_\mathrm{sc}$ becomes field-independent. By analysing the temperature of the magnetic penetration depth, within the local London approximation, a direct association with the superconducting gap symmetry can be made \cite{musrfit}:
\begin{equation}
    \frac{\lambda^{-2}(T, \Delta_{0,i})}{\lambda^{-2}(0, \Delta_{0,i})} = 1 + \frac{1}{\pi} \int_0^{2\pi} \int_{\Delta(T, \phi)}^\infty \left(\frac{\delta f}{\delta E} \right) \frac{EdEd\varphi}{\sqrt{E^2-\Delta_i(T,\varphi)^2}}
\end{equation}
where $f = [1+\mathrm{exp}(E/k_BT)]^{-1}$ is the Fermi function, $\varphi$ is the angle along the Fermi surface, and $\Delta_i(T, \varphi) = \Delta_{0,i}\Gamma(T/T_\mathrm{C})g(\varphi)$ ($\Delta_{0,i}$ is the maximum gap value at $T = 0$). The temperature dependence of the gap is approximated by the expression, $\Gamma(T/T_\mathrm{C} = \mathrm{tanh}{1.82[1.018(T_\mathrm{C}/T - 1)]^{0.51}}$ \cite{sc_gap}, whilst $g(\varphi)$ describes the angular dependence of the new gap and is replaced by $1$ for an $s$-wave gap, $[1+a\mathrm{cos}(4\varphi)/(1 + a)]$ for an anisotropic $s$-wave gap, and $|\mathrm{cos}(2\varphi)|$ for a $d$-wave gap \cite{sc_gap2}.\\

\section{Acknowledgments}~
The ${\mu}$SR experiments were carried out at the Swiss Muon Source (S${\mu}$S) Paul Scherrer Insitute, Villigen, Switzerland. Z.G. acknowledges support from the Swiss National Science Foundation (SNSF) through SNSF Starting Grant (No. TMSGI2${\_}$211750). Z.G. acknowledges the useful discussions with Dr. Pietro Bonfa, Prof. Samuele Sanna, Prof. Stephen Wilson and Dr. Robert Scheuermann. The work at BIT was supported by the National Key R${\&}$D Program of China (No. 2020YFA0308800, and No. 2022YFA1403400), the National Natural Science Foundation of China (No. 92065109),the Beijing National Laboratory for Condensed Matter Physics (Grant No. 2023BNLCMPKF007), the Beijing Natural Science Foundation (Grant No. Z210006). Z.W. thanks the Analysis ${\&}$ Testing Center at BIT for assistance in facility support.\\

\section{Author contributions}~
Z.G. conceived and supervised the project. 
Crystal growth: Y.L., Y.Y., and Z.W.. 
${\mu}$SR and AC susceptibility experiments as well as the corresponding discussions: J.N.G., S.S.I., V.S., G.J., P.K., O.G., A.D., I.B., J.C., Z.S., A.S., T.P., H.L., R.K., and Z.G..  
${\mu}$SR data analysis: J.N.G. and Z.G..
STM experiments and corresponding discussions: H.D., J.N.G., Z.G., and J.-X.Y.. ARPES experiments: Y.Z. and K.O.. Figure development and writing of the paper: J.N.G. and Z.G. with contributions from all authors. All authors discussed the results, interpretation, and conclusion.\\

\subsection*{Data availability}
All related data are available from the authors. Alternatively, all the muon-spin rotation data can be accessed through the SciCat data base using the following links:
GPS: \href{http://doi.psi.ch/detail/10.16907\%2F83b854b1-00fd-43e2-9d25-fa5d2d2eac07}{doi.psi.ch/detail/10.16907/83b854b1-00fd-43e2-9d25-fa5d2d2eac07} 
LEM: \href{http://doi.psi.ch/detail/10.16907\%2F1c2661e3-769b-4cdf-8504-4959fff59d17}{doi.psi.ch/detail/10.16907/1c2661e3-769b-4cdf-8504-4959fff59d17} , and other data at \href{http://musruser.psi.ch/cgi-bin/SearchDB.cgi}{http://musruser.psi.ch/cgi-bin/SearchDB.cgi} using the following details:  1. 1. Area: Dolly, Year: 2023, Run Title: CVS-Rb..., Run: 0263 - 0291 (named incorrectly in file). 2. Area: HAL, Year: 2023, Run Title: CVS..., Run: 0047 - 0182. 3. Area: GPD, Year: 2024, Run Title: Nb-CVS..., Run: 0067 - 0181.

\bibliography{References}{} 

\section*{Extended Data}
\subsection*{AC Susceptibility data}
\begin{figure*}
    \centering
    \includegraphics[width=0.5\textwidth]{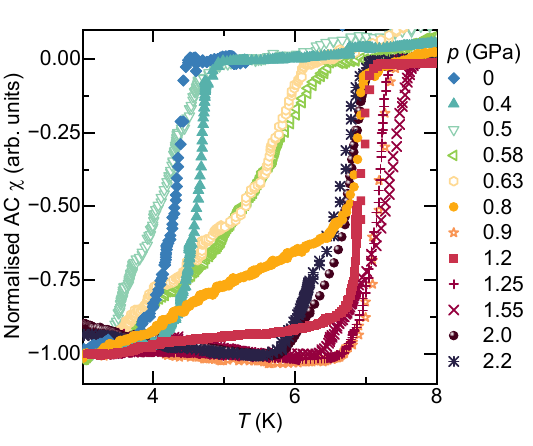}
    \caption{\textbf{Full AC Susceptibility data used to create the pressure phase diagram in Fig. \ref{fig_AC_pressure}a.}
    Solid markers are the same susceptibility curves shown in Fig. \ref{fig_musr_pressure}a, and open markers are additional susceptibility measurements.}
    \label{Ext_data1}
\end{figure*}

The full AC susceptibility data set used to create the pressure phase diagram (Fig. \ref{fig_AC_pressure}a) is shown in Fig. \ref{Ext_data1}. The data was collected using an inhouse Janis cryostat system, with the temperature controlled by regulating the flow of liquid helium. The pressure was determined by tracking the superconducting transition of indium against a reference indium outside the pressure cell---both were removed from the susceptibility curves for clarity. Then to correct for differences in the coil position, each curve was normalised between $-1$ and $0$. The normalisation points were chosen when the transition had flattened out, or in instances of large background contributions, directly preceding or following the transition. The phase diagram was constructed by plotting each of the susceptibility curves together in a $3$D colour plot. For clarity, data with a higher AC value of zero were artificially flattened in order to make the transition clear.

\subsection*{ARPES characterization of the charge order gap at M point}
To compare the charge order gap, we conducted angle-resolve photoemission spectroscopy (ARPES) measurements on pristine and 7$\%$ Nb-doped CsV$_{3}$Sb$_{5}$ samples. We used a photon energy of 21.218 eV obtained from a Helium discharging lamp (Scienta Omicron VUV 5050) and captured photoelectrons with a Scienta R4000 analyzer. The single-crystal samples were cleaved in-situ and measured under vacuum conditions better than 3$\times10^{-11}$ torr, with an energy resolution set to 15 meV.  Figure 7 presents the ARPES spectra of both pristine and 7$\%$-Nb doped CsV$_{3}$Sb$_{5}$ along the KMK momentum cut. To enhance the visibility of the gap near the Fermi level ($E_{\rm F}$), we divided the spectra by the Fermi-Dirac function. In the pristine sample spectra shown in Fig. 7(a) and (b), a clear gap appears around the M point (k// = 0) in the charge ordered state, which closes in the normal state. With 7$\%$-Nb doping, which partially suppresses the charge order transition, the gap near $E_{\rm F}$ remains but becomes less blurred, as illustrated in Fig. 7(c). Figure 7(e) shows the energy distribution curves (EDCs) at the M point, where the charge order gap is indicated by a leading-edge shift in the EDC of the charge ordered state compared to the one of the normal state. To directly compare the charge order gap between these two samples, in Fig. 7(f), we symmetrized the EDCs taken in the charge ordered state at $T$ =  7 K. While the gap, determined from the peak in the symmetrized EDCs, remains around 24 meV for both samples, the 7$\%$-Nb doping results in a significant filling of the gap, highlighting the partial suppression of charge order in the doped sample.

\begin{figure*}
    \centering
    \includegraphics[width=\textwidth]{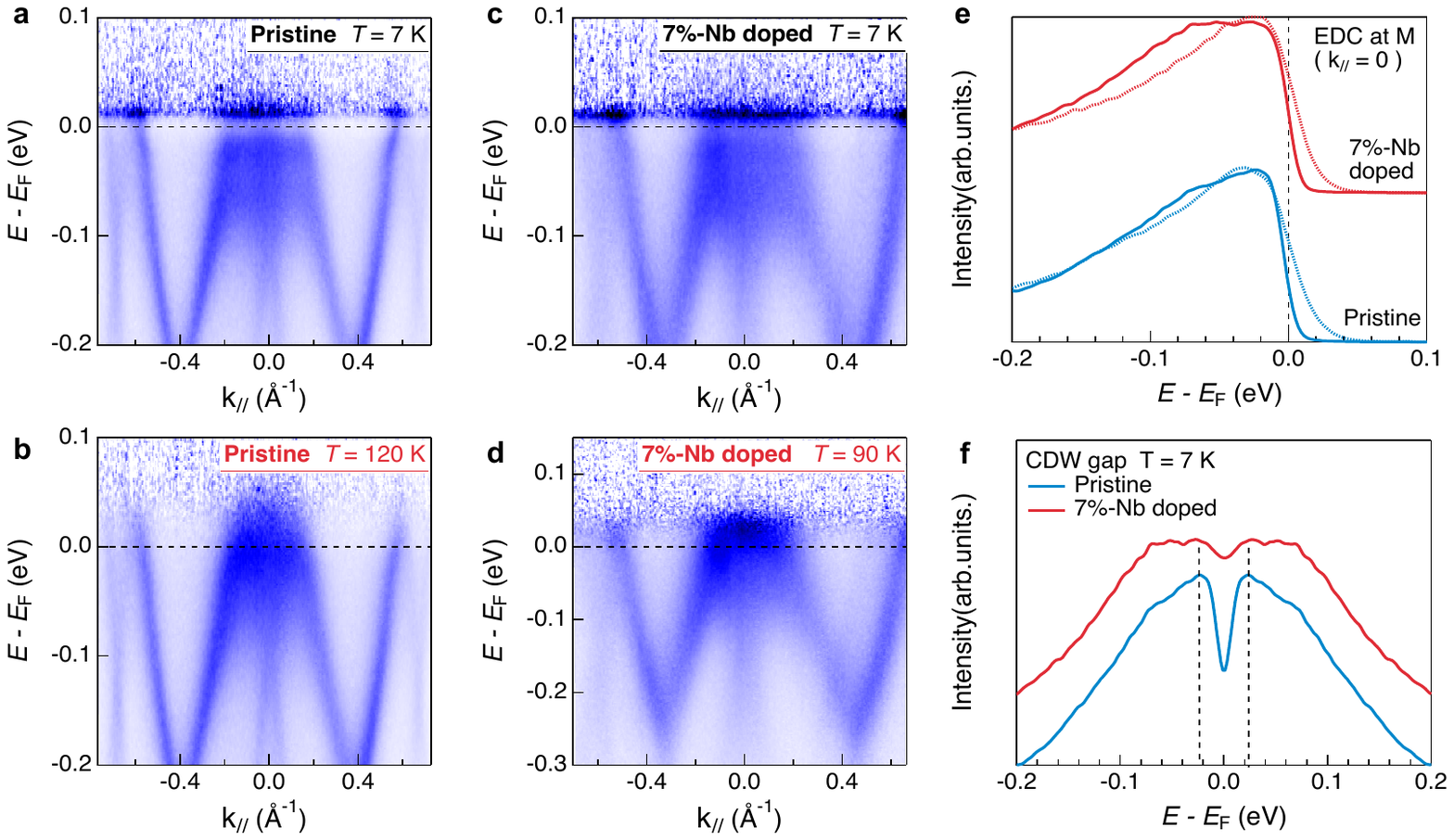}
    \vspace{-2.0cm}
    \caption{\textbf{ARPES measurements.} 
    \textbf{a} Band dispersion of pristine CsV$_{3}$Sb$_{5}$ along the KMK cut in the charge ordered state at $T$ = 7 K, and \textbf{b} in the normal state at $T$ = 120 K. \textbf{c} Band dispersion of 7$\%$-Nb doped CsV$_{3}$Sb$_{5}$ along the KMK cut in the charge ordered state at $T$ = 7 K, and \textbf{d} in the normal state at $T$ = 90 K. The spectra were divided by the Fermi-Dirac function to better visualise the gap near $E_{\rm F}$. \textbf{e} EDCs at the M point (k// = 0) with the solid lines representing the charge ordered state and the dashed lines representing the normal state. \textbf{f} Symmetrised EDCs at the M point in the charge ordered state.}
    \label{fig_arpes}
\end{figure*}

\end{document}